\documentclass[letterpaper,twocolumn,prl,
aps,showpacs,superscriptaddress,floatfix]{revtex4-2}

\usepackage[latin1]{inputenc}
\usepackage{bbm}
\usepackage{bm}
\usepackage[usenames]{color}
\usepackage{multirow}
\usepackage{amssymb}
\usepackage{amsbsy}
\usepackage{mathtools}
\usepackage{amsmath}
\usepackage{stmaryrd}
\usepackage{graphicx}
\usepackage{epsfig}
\usepackage{placeins}
\usepackage{bbold}
\usepackage{braket}
\usepackage{blindtext}
\usepackage[colorlinks,linkcolor=blue,citecolor=blue,urlcolor=blue]{hyperref}

\usepackage{scalerel}
\usepackage{tikz}
\usetikzlibrary{calc}
\usetikzlibrary{patterns}
\usetikzlibrary{svg.path}
\definecolor{orcidlogocol}{HTML}{A6CE39}
\tikzset{
	orcidlogo/.pic={
		\fill[orcidlogocol] svg{M256,128c0,70.7-57.3,128-128,128C57.3,256,0,198.7,0,128C0,57.3,57.3,0,128,0C198.7,0,256,57.3,256,128z};
		\fill[white] svg{M86.3,186.2H70.9V79.1h15.4v48.4V186.2z}
		svg{M108.9,79.1h41.6c39.6,0,57,28.3,57,53.6c0,27.5-21.5,53.6-56.8,53.6h-41.8V79.1z M124.3,172.4h24.5c34.9,0,42.9-26.5,42.9-39.7c0-21.5-13.7-39.7-43.7-39.7h-23.7V172.4z}
		svg{M88.7,56.8c0,5.5-4.5,10.1-10.1,10.1c-5.6,0-10.1-4.6-10.1-10.1c0-5.6,4.5-10.1,10.1-10.1C84.2,46.7,88.7,51.3,88.7,56.8z};
	}
}

\newcommand\orcid[1]{\!%
  \href{https://orcid.org/#1}{%
    \mbox{%
      \scaleto{%
        \begin{tikzpicture}[yscale=-1,transform shape]
          \pic{orcidlogo};
        \end{tikzpicture}
      }{8pt}%
    }%
  }%
}

\makeatletter

\newcommand{\be}{\begin{equation}}
\newcommand{\ee}{\end{equation}}
\newcommand{\new}[1]{\textcolor{black}{#1}}

\makeatletter

\begin{document}
\title{
Estimate of equilibration times of quantum correlation functions in the thermodynamic limit based on Lanczos coefficients
}
 \date{\today}
\author{Jiaozi Wang~\orcid{0000-0001-6308-1950}}
\affiliation{Department of Mathematics/Computer Science/Physics, University of Osnabr\"uck, D-49076 
Osnabr\"uck, Germany}
\author{Merlin F\"{u}llgraf~\orcid{0009-0000-0409-5172}}
\affiliation{Department of Mathematics/Computer Science/Physics, University of Osnabr\"uck, D-49076 
Osnabr\"uck, Germany}
\author{Jochen Gemmer~\orcid{0000-0002-4264-8548}}
\affiliation{Department of Mathematics/Computer Science/Physics, University of Osnabr\"uck, D-49076 
Osnabr\"uck, Germany}

\begin{abstract}
We study the equilibration times $T_\text{eq}$ 
of local observables in quantum chaotic systems by considering their autocorrelation functions.
Based on the recursion method, we suggest a scheme to estimate $T_\text{eq}$ from the corresponding Lanczos coefficients that is expected to hold in the thermodynamic limit.
We numerically find that if the observable eventually shows smoothly  growing Lanczos coefficients, a finite number of the former is sufficient for a reasonable estimate of the equilibration time.
This implies that equilibration occurs on a realistic time scale much shorter than the life of the universe. The numerical findings are further supported by analytical arguments.

\end{abstract}
\maketitle
{\it Introduction.\ }
Whether and how a quantum many-body system approaches equilibrium has long been an important question. It has attracted great attention, both theoretically \cite{Gogolin_2016-review,PhysRevX.7.031027-ET-Short17,goldstein2015extremely-ET-Goldstein15,PhysRevLett.106.130601-ET-Kastner11,PhysRevE.110.024126-ET-Christian24,reimann2016typical-ET-Reimann16,PhysRevE.90.012121-ET-Short14,PhysRevLett.123.200604-ET-Eisert19,PhysRevLett.111.140401-ET-Goldstein13,PhysRevLett.124.110605_ET_Pintos20,Gogolin_2016-review} and experimentally \cite{geiger2014local-ET-Exp-Tim14,langen2013local-ET-Exp-Tim13,richerme2014non-ET-Exp-Monroe14,trotzky2012probing-ET-Exp-Bloch12,jurcevic2014quasiparticle-ET-Exp-Zoller14}. 

Over the past few decades, substantial progress has been made, thanks to the (re)introduction and development of concepts like typicality and the eigenstate thermalization hypothesis (ETH). 
\new{The ETH explains eventual thermalization of non-integrable systems after sufficiently long times by
postulating a particular structure of matrix elements of
the observable in the eigenbasis of the Hamiltonian\cite{ETH-PhysRevA.43.2046,ETH-PhysRevE.50.888,ETH-rigol2008thermalization}.
However, it fails to address the question why for physical systems equilibration happens on a conceivable time scale, much shorter than the age of the universe--a question that remains open and actively debated.}
Analytical efforts have primarily focused on establishing bounds for equilibration time. In recent years, several bounds have been rigorously derived \cite{Short_2012-ET,Reimann_2012-ET,PhysRevE.102.032144-ET,PhysRevX.7.031027-ET-Short17,PhysRevE.90.012121-ET-Short14,PhysRevLett.124.110605-ET,de_Oliveira_2018},  significantly improving our understanding of the timescale of equilibration.
However, in the existing analytical studies, the general assumptions underlying the approaches are often violated in typical physical settings, see e.g.\ Refs.\ \cite{Heveling_2020,heveling_PhysRevX.10.028001}.
Consequently, questions arise considering the relevance of such bounds to the equilibration in real systems.

In addition to the analytical work, the problem has also been studied numerically \cite{Genway12-PhysRevA.86.023609}.
Due to limitations on the system size that are numerically accessible, the equilibration timescale in the thermodynamic limit remains an open question.



 
In our recent work \cite{PhysRevE.110.024126-ET-Christian24}, we tackled this problem employing an approach based on the so-called recursion method. In this work we elaborate on a direct scheme, based on \new{smoothly} growing Lanczos coefficients, to infer the equilibration time in the thermodynamic limit. Numerically, we find that if the Lanczos coefficients exhibit such a behaviour, our method converges quickly, indicating a reasonable and cheap estimate, once the first several Lanczos coefficients are taken into account. Furthermore, we present analytical arguments supporting our numerical findings and suggesting the typicality of finite equilibration times in real physical systems.
 

{\it Framework.\ }
 Given an observable $\cal O$ and a Hamiltonian $H$ 
we are interested in the autocorrelation
\begin{equation}
{\cal C}(t)=\frac{\text{Tr}[{\cal O}(t){\cal O}]}{\text{Tr}[{\cal O}^{2}]},
\end{equation}
where ${\cal O}(t)=e^{iHt}{\cal O}e^{-iHt}$. In our paper, we focus on the following definition of equilibration time \cite{PhysRevLett.124.110605_ET_Pintos20}
\begin{equation}
T_{\text{eq}}:=\int_{0}^{\infty}|{\cal C}(t)-{\cal C}(\infty)|^{2}dt,
\end{equation}
where
$\mathcal{C}(\infty)=\lim_{T\rightarrow\infty}\frac{1}{T}\int_{0}^{T}\mathcal{C}(t)dt$.
\new{Assuming $\mathcal{C}(\infty) = 0$, $T_\text{eq}$ becomes}
\begin{equation}\label{eq_def_teq}
    T_{\text{eq}}=\int_{0}^{\infty}{\cal C}^2(t)dt.
\end{equation}
The equilibration time $T_\text{eq}$ can be interpreted as the area under the curve of ${\cal C}^2(t)$. 
\new{To illustrate the physical meaning of $T_\text{eq}$ defined above, consider the simple example ${\cal C}(t) = e^{-\Gamma t} \cos(\omega t)$, for which
$T_\text{eq} = \frac{1}{4\Gamma} + \frac{\Gamma}{4(\Gamma^{2} + \omega^{2})} \sim \frac{1}{\Gamma}.$
Without the square in Eq.~(\ref{eq_def_teq}), the integral
$\int_{0}^{\infty} {\cal C}(t) \, dt = \frac{\Gamma}{\Gamma^{2} + \omega^{2}}$
would fail to capture the relevant timescale of thermalization, e.g.,  in case of rapid oscillation $\omega\gg \Gamma$.}



\new{To introduce the general framework}, it is convenient to switch to the
Hilbert of space of operators, also Liouville space, and denote its elements $\cal O$ by states $|{\cal O})$. This
space is equipped with an inner product $({\cal O}_m|{\cal O}_n) = \text{Tr}[{\cal O}^\dagger_m
{\cal O}_n]/\text{Tr}[\mathbb{1}]$, which defines a norm via $\Vert {\cal O} \Vert = \sqrt{({\cal O}|{\cal O})}$. The Liouvillian
superoperator is defined by \new{${\cal L}|{\cal O}) =\vert [H, {\cal O}])$}, where $H$ denotes the Hamiltonian of the system, and propagates a state $|{\cal O})$ in time, such that an autocorrelation function (at infinite temperature) can be written as 
\begin{align}
    {\cal C}(t) = ({\cal O}|e^{i{\cal L}t}|{\cal O}).\label{eq_autocorrelation}
\end{align}

Starting from an initial ``seed" ${\cal O}$, a \new{tridiagonal} representation of
$\cal L$ can be obtained using the Lanczos algorithm. From a normalized initial state
$|{\cal O}_0) \propto |{\cal O})$, i.e., $({\cal O}_0|{\cal O}_0)=1$, we set $b_1 = \Vert {\cal L}\vert{\cal O}_0)
\Vert$ as well as $|{\cal O}_1) = {\cal L} |{\cal O}_0)/b_1$. Then, we iteratively compute
\begin{align}
    \begin{split}
        |\widetilde{\cal O}_n) &= {\cal L} |{\cal O}_{n-1}) - b_{n-1} |{\cal O}_{n-2}) \, , \\
b_n &= \Vert \widetilde{\cal O}_n \Vert \, ,  \\
 |{\cal O}_n) &= |\widetilde{\cal O}_n)/b_n \, . \label{eq-Lanczos}
    \end{split}
\end{align}
The tridiagonal representation of $\cal L$ in the Krylov basis
$\{ |{\cal O}_n) \}$ results as
\begin{equation}
{\cal L}_{mn} = ({\cal O}_{m}| {\cal L} |{\cal O}_{n}) = 
\delta_{m,n+1}b_n+\delta_{m,n-1}b_m,
\end{equation}
where the coefficients $b_{n}$ are real and positive numbers, referred to as Lanczos coefficients. The (infinite) set of Lanczos coefficients uniquely determines the autocorrelation function and vice versa. Their
iterative computation is an elementary part of the recursion method. Note that it is possible to calculate a certain number of  $b_n$ (practically in the lower two digit regime) even for infinitely large systems, i.e., in the thermodynamic limit, if the Hamiltonian and the observable are local. In the remainder of the paper at hand we address this scenario.
Furthermore, in Mori formalism\ \cite{Gray86-rc,10.1143/PTP.33.423-Mori}, the Lanczos coefficients $b_n$ are directly related to the Laplace transform of the respective autocorrelation function via a continued fraction representation of form
\begin{equation}
        {\cal F}(s)=\int_{0}^{\infty}e^{ts}{\cal C}(t)dt=\frac{1}{s+\frac{b_{1}^{2}}{s+\frac{b_{2}^{2}}{s+\frac{b_{3}^{2}}{\cdots}}}}\label{eq_mori_all_c},
\end{equation}
see also \cite{wang2023diffusion-RM-wang24}.
\new{From Eq.\ (\ref{eq_mori_all_c}), we infer that the infinite-time integral of $\mathcal{C}(t)$ is solely determined by the Lanczos coefficients as}
\begin{align}
    \mathcal{F}(0)=\int_{0}^\infty \mathcal{C}(t)dt=\frac{1}{b_{1}}\prod_{n=1}^{\infty}\left(\frac{b_{n}}{b_{n+1}}\right)^{(-1)^{n}}.\label{eq-int-c}
\end{align}
Consequently, returning to the original question, the equilibration time $T_{\text{eq}}$ defined in Eq.\ (\ref{eq_def_teq}) may computed by virtue of the Lanczos coefficients related to the autocorrelation function $\mathcal{C}^2(t)$, denoted by $B_n$.


To study \new{the} $B_N$, we consider a product space spanned by $\{|mn)\} :=\{|{\cal O}_m)\otimes|{\cal O}_n)\}$, and a Liouvillian superoperator ${\cal L}^\prime = {\cal L}\otimes \mathbb{1} + \mathbb{1} \otimes {\cal L}$. In this setting the observable corresponding to $\mathcal{C}^2(t)$ may be understood as $\mathcal{O}^\prime=\mathcal{O}\otimes\mathcal{O}$, such that
\begin{align}
    \mathcal{C}^2(t)=(\mathcal{O}^\prime\vert e^{i\mathcal{L}^\prime t}\vert\mathcal{O}^\prime)=({\cal O}|e^{i{\cal L}t}|{\cal O})({\cal O}|e^{i{\cal L}t}|{\cal O}),
\end{align}
 relating the setting to the simple case as in Eq.\ (\ref{eq_autocorrelation}).
\new{The Lanczos coefficient $B_N$ is calculated using ${\cal L}^\prime$ and a ``seed" operator in the product space $|{\cal Q}_0)$ = $|00)$. Specifically, we determine $B_1=\Vert\mathcal{L}^\prime\vert\mathcal{Q}_0)\Vert$ and set $\vert\widetilde{\mathcal{Q}}_1)=\mathcal{L}^\prime\vert\mathcal{Q}_0)$ and again follow the Lanczos scheme}
\begin{align}
    \begin{split}
        |\widetilde{{\cal Q}}_{N})&={\cal L}^{\prime}|{\cal Q}_{N-1})-B_{N-1}|{\cal Q}_{N-2}),\\
B_N&=\Vert \widetilde{\mathcal{Q}}_N\Vert,\\
\vert \mathcal{Q}_N)&=\vert\widetilde{\mathcal{Q}}_N)/B_N.
\label{eq-Bn}
    \end{split}
\end{align}
Given \new{the} $b_n$, \new{the} $B_N$ can be calculated straightforwardly using Eq.~\eqref{eq-Bn}, e.g.,
\begin{equation}\label{eq-Bbn}
  \begin{split}
      &\vert\widetilde{\mathcal{Q}}_1) =b_1(\vert 10) + \vert 01)),  \\ &B_1 = \sqrt{2}b_1,\ \ldots, B_2 =\sqrt{2b_{1}^{2}+b_{2}^{2}},\ \ldots ,
  \end{split}
\end{equation}
which indicates that $B_N$ are unambiguously determined by 
$b_{n(n \le N)}$.
With the Lanczos coefficients $B_N$ we can eventually formulate the equilibration time $T_{\text{eq}}$ similarly to Eq.\ (\ref{eq-int-c}) as
\begin{equation}\label{eq-eqtime-Bn}
T_{\text{eq}}\equiv\int_0^\infty \mathcal{C}^2(t)dt=\frac{1}{B_{1}}\prod_{N=1}^{\infty}\left(\frac{B_{N}}{B_{N+1}}\right)^{(-1)^{N}}.
\end{equation}
 Crucially, the scheme described in Eq.\ (\ref{eq-Bn}) is considerably simpler than the original Lanczos algorithm for $\mathcal{C}(t)$, Eq.\ (\ref{eq-Lanczos}), once by virtue of the coefficients $b_n$ the operators $\mathcal{L}_{mn}$ are known. This is our first main result of the paper.

In practice, only the first several $b_n$ are easily numerically accessible. Let  $n_{\text{max}}$ denotes the number of feasible $b_n$. \new{From these $b_n$ we calculate the first $B_{N(N \leq  R)}$ using the scheme (\ref{eq-Bn}), which is applicable up to $R\le n_\text{max}$. }
\new{For convenience in the following discussion, let us rewrite Eq.~\eqref{eq-eqtime-Bn} as
\begin{equation}\label{eq-Teq0}
    T_{\text{eq}}=\begin{cases}
\frac{1}{p_{R}B_{R}}\prod_{M=1}^{\frac{R}{2}}\frac{B_{2M}^{2}}{B_{2M-1}^{2}} & \text{, even }R\\
\frac{p_{R}}{B_{R}}\prod_{M=1}^{\frac{R-1}{2}}\frac{B_{2M}^{2}}{B_{2M-1}^{2}} & \text{, odd }R
\end{cases},
\end{equation}
where
\begin{equation}
p_{R+1}=\!\!\prod_{M=1}^{\infty}\!\!\left(\frac{B_{R+M}}{B_{R+M+1}}\right)^{(-1)^{M}} .
\end{equation}
Now we resort to an assumption that the remaining  $B_{N(N>R)}$ grow linearly, $B_N(N>R) = \alpha_R N + \beta_R$. In this case, $p_R$ 
can be derived analytically (detailed in supplemental material),
\begin{equation}\label{eq-pn-origin}
\tilde{p}_{R}=\frac{\Gamma(\frac{R}{2}+\frac{\beta_{R}}{2\alpha_{R}})\Gamma(\frac{R}{2}+\frac{\beta_{R}}{2\alpha_{R}}+1)}{\Gamma^{2}(\frac{R}{2}+\frac{\beta_{R}}{2\alpha_{R}}+\frac{1}{2})}\ .
\end{equation}
Here, we use a rather simple approach, where $\alpha_R$ and $\beta_R$ are obtained from a linear extrapolation of the two last exact $B_N$, i.e.\ , $\alpha_R =  B_{R} - B_{R-1}$ and $\beta_R=R B_{R-1}-(R-1)B_R$. 
This yields the following estimate:
\begin{equation}\label{eq-T-rc}
T_{\text{eq}}\simeq T_{\text{eq}}^{\text{rc}}= \begin{cases}
\frac{1}{\tilde{p}_{R}B_{R}}\prod_{M=1}^{\frac{R}{2}}\frac{B_{2M}^{2}}{B_{2M-1}^{2}} & \text{, even }R\\
\frac{\tilde{p}_{R}}{B_{R}}\prod_{M=1}^{\frac{R-1}{2}}\frac{B_{2M}^{2}}{B_{2M-1}^{2}} & \text{, odd }R
\end{cases}.
\end{equation}}
This estimate would be exact if the above extrapolation captured the true $B_N$ precisely. \new{More importantly, it may be still very close, even if the $B_{N(N>R)}$ are not precisely linear but {smooth} (for a formal definition see Eq. (\ref{eq-smooth})).} This is tantamount to stating that  $T_{\text{eq}}$ is not very sensitive to the $B_N$ at larger $N$ as long as they are smooth.  Fig.~S7 in the supplemental material provides an instructive example for $b_N \propto \sqrt{N}$.  

\new{At this point a crucial conceptual question arises: Under which conditions will the $B_N$ become eventually smooth? If they do so, this entails a technical question: Is the practically available number of $b_n$ large enough such that $R$ reaches into a regime in which the true $B_N$ are sufficiently smooth? }

To address the conceptual question, here we rely on an argument  which is presented in detail in the supplemental material: Given a \textit{smooth} profile of the original $b_n$ above some $n_s$, we approximately find 
\begin{equation} \label{Bb}
B_N\approx2b_{N/2}
\end{equation}
for $N\ge2n_s$. This means the $B_N$ may  inherit smoothness from the $b_n$.  Smoothness of the $b_n$ may, however, be to some extent based on the \textit{operator growth hypothesis} \cite{Parker19}. It states that in infinite chaotic quantum many-body systems the Lanczos coefficients of local operators \textcolor{black}{that have no overlap with any conserved quantity,} are asymptotically linear (with logarithmic corrections in one dimension).  Thus, as the convergence of $  T_{\text{eq}}^{\text{rc}}   $ occurs for smooth $b_n$, the former is generally expected for \textcolor{black}{such} local observables in chaotic quantum systems.  \new{The answer to the technical question, however, will rely more on the numerical results presented below.}

\begin{figure}[t]
	\centering
    \includegraphics[width = 1.0\linewidth]{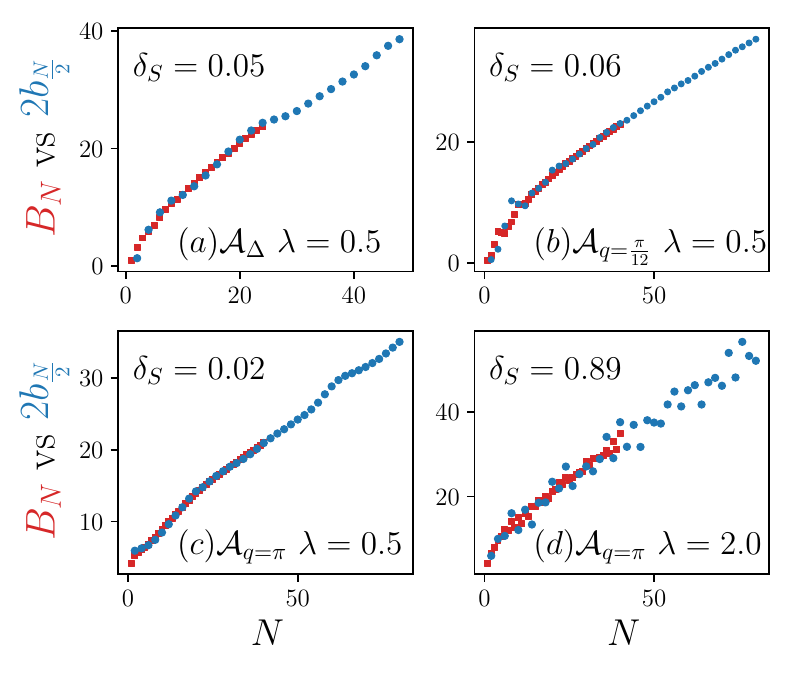}
	\caption{\new{Lanczos coefficients $B_N$ (red square) related to the autocorrelation function $\mathcal{C}^2$ versus $2 b_{N/2}$ (blue dot) with $b_n$ the Lanczos coefficients related to $\mathcal{C}$, for different models, observables and system parameters.
  (a): Ising ladder, energy difference operator ${\cal A}_\Delta$, $\lambda = 0.5$; and tilted field Ising model, energy density wave operator ${\cal A}_{q}$ for (b): $q = \pi/12$, $\lambda = 0.5$; (c): $q = \pi$, $\lambda = 0.5$ and (d): $q= \pi$,  $\lambda = 2.0$.
    $q$ indicates the wavenumber of the density-wave operator ${\cal A}_q$.
    $\delta_S$ is the smoothness indicator defined in Eq.~\eqref{eq-smooth}.}
    }\label{Fig1}
\end{figure}


{\it Numerical results.\ }
\new{To check our assumptions and main results, as well as to address the technical question stated above}, we consider some physical models for which we compute $n_{\max}$ Lanczos coefficients per operator. The number $n_{\max}$ differs for each model because of the numerical complexity. However, all $n_{\max}$ Lanczos coefficients correspond to the respective infinite system.
The first model we consider is an Ising ladder, 
\begin{equation}
    H = H_1 + \lambda H_I + H_2,
\end{equation}
where
\begin{gather}
    H_{r}=\sum_{\ell=1}^{L/2}h_{x}\sigma_{r,\ell}^{x}+h_{z}\sigma_{r,\ell}^{z}+J\sigma_{r,\ell}^{z}\sigma_{r,\ell+1}^{z}, \nonumber \\
    H_{I}=\sum_{\ell=1}^{L/2}\sigma_{1,\ell}^{z}\sigma_{2,\ell}^{z} .
\end{gather}
Here $\sigma_{r,\ell+1}^{x,z}$ indicates the Pauli matrix at site $(r, \ell)$, \textcolor{black}{where $r=1,2$ and $\ell = 1,2,\ldots,L/2$. Here $L$ is chosen to be even}.
The operator of interest is the energy-difference operator
${\cal A}_\Delta = H_1 - H_2$. Parameters are chosen as $h_x = 1.0,\ h_z = 0.5,\ J = 1.0$. For this model we obtain \new{$n_{\max}=24$} Lanczos coefficients.

\begin{figure}[t]
	\centering
    \includegraphics[width = 1.0\linewidth]{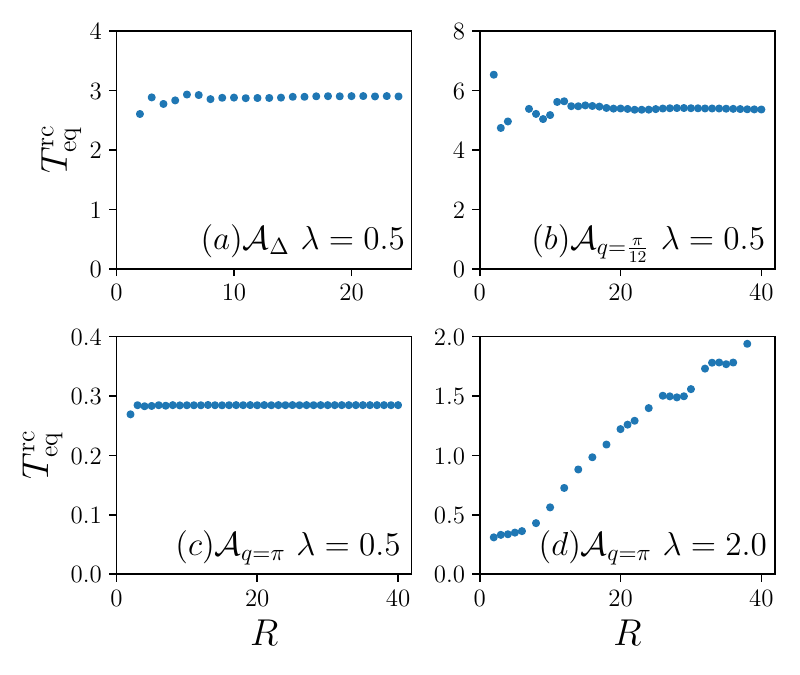}
	\caption{\textcolor{black}{Equilibration time} $T^{\text{rc}}_{\text{eq}}$ (given in Eq.~\eqref{eq-T-rc}) \textcolor{black}{for different numbers} $R$ \textcolor{black}{of included Lanczos coefficients}, for \textcolor{black}{different models, observables and system parameters.} (a): Ising ladder, energy difference operator ${\cal A}_\Delta$, $\lambda = 0.5$; and tilted field Ising model, energy density wave operator ${\cal A}_{q}$ for (b):$q = \pi/12$, $\lambda = 0.5$; (c):$q = \pi$, $\lambda = 0.5$ and (d): $q= \pi$,  $\lambda = 2.0$.
}\label{Fig2}
\end{figure}

As a second model, we study a tilted field Ising (TFI) chain,
\begin{equation}
   H=\sum_{\ell=1}^{L}h_{x}\sigma_{\ell}^{x}+\lambda\sigma_{\ell}^{z}+J\sigma_{\ell}^{z}\sigma_{\ell+1}^{z},
\end{equation}
where we focus on energy density wave operators ${\cal A}_{q}=\sum_{\ell=1}^{L}\cos(q\ell)h_{\ell}$, where
$h_{\ell}$ indicates the local energy 
\begin{equation}
    h_{\ell}=\frac{h_{x}}{2}(\sigma_{\ell}^{x}+\sigma_{\ell+1}^{x})+\frac{\lambda}{2}(\sigma_{\ell}^{z}+\sigma_{\ell+1}^{z})+J\sigma_{\ell}^{z}\sigma_{\ell+1}^{z}.
\end{equation}
In the numerical investigation,  we fix $h_x=1.05,\ J = 1.0$.  Two different wave numbers are considered here, $q=\pi$ and $q=\pi/12$. Here, for both observables we obtain \new{$n_{\max}=40$} Lanczos coefficients.

As a start, we show some examples of Lanczos coefficients in Fig.~\ref{Fig1}. To characterize the smoothness of $b_n$, we introduce a ``smoothened" version of the $b_n$, given by $\tilde{b}_{n}:=\frac{1}{4}b_{n-1}+\frac{1}{2}b_{n}+\frac{1}{4}b_{n+1}\  (n\ge 2)$.  The smoothness $\delta_S$ is thus defined as
\begin{equation}\label{eq-smooth}
\delta_{S}=\frac{1}{n_{\text{max}}-2}\sum_{n=2}^{n_{\text{max}} - 1}|b_{n}-\tilde{b}_{n}|.
\end{equation}
In [(a)(b)(c)], approximately smooth $b_n$ with a small $\delta_S$ are observed, whereas in (d), significant fluctuations in $b_n$ are evident, corresponding to a large $\delta_S$, see Fig.\ \ref{Fig1}.

Moreover, we verify $B_{N} \approx 2b_{\frac{N}{2}}$, \new{where $B_N$ is computed from the scheme (\ref{eq-Bn}).} When the original $b_n$ become smooth for growing $n$ (as indicated by a small $\delta_S$), there is a good agreement, as shown in [(a)(b)(c)]. In contrast, if the $b_n$ are not smooth, their non-smooth structure carries over to \new{the} $B_N$, \new{leading to deviations between 
$B_N$ and $2 b_{\frac{N}{2}}$}. This is in accord with the principles stated around Eq. (\ref{Bb}), and detailed in the supplemental material. 

In Fig.\ \ref{Fig2} we exemplary show our estimation of $T_{\text{eq}}^{\text{rc}}$ for varying $R$ for the cases [(a)-(d)] discussed in Fig.\ \ref{Fig1}. In the cases of \textit{smooth} Lanczos coefficients, here [(a)-(c)] as indicated by small $\delta_S$, $T_{\text{eq}}^{\text{rc}}$ only varies slightly with respect to $R$, for $R\gtrsim5$. This observation suggests that the first several Lanczos coefficients are sufficient to obtain a reasonable estimate of $T_{\text{eq}}$.
Conversely, in the scenario (d), where the Lanczos coefficients are not smooth (large $\delta_S$), the quantity $T_{\text{eq}}^{\text{rc}}$ does not converge, indicating that an accurate result for $T_{\text{eq}}$ might not yet have been reached.

Strictly speaking, an apparent convergence of $T_{\text{eq}}^{\text{rc}}$ at finite $R$ does not yet warrant the correctness of the estimate. Thus  we compare $T_{\text{eq}}^{\text{rc}}$ with 
results from direct simulations of the autocorrelation function. More precisely, we study the pendant of the equilibration time, see Eq.\ (\ref{eq_def_teq}), given by $ T^\text{typ}_\text{eq} := \int_0 ^{T_c} {\cal C}^2(t) dt$, with appropriate $T_c$, for details see \cite{T-Typ}.
 Here we consider a finite system size $L = 24$, where ${\cal C}(t)$ is calculated using dynamical quantum typicality \cite{PhysRevLett.102.110403-Christian-DQT}. (Non-systematic checking indicates that for all considered cases         $ T^\text{typ}_\text{eq}$ does not strongly depend on $L$ for $L\geq 24$.)

\begin{figure}[t]
	\centering
    \includegraphics[width = 0.9\linewidth]{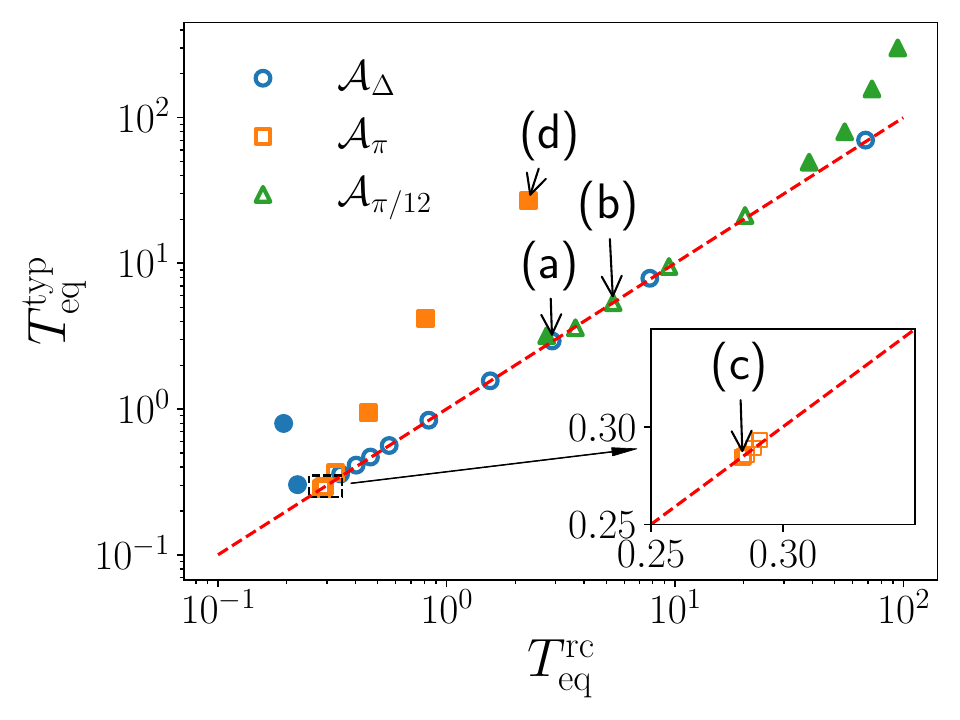}
	\caption{Comparison between the equilibration time estimated using \textcolor{black}{the} recursion method $T^\text{rc}_\text{eq}$ (Eq.~\eqref{eq-T-rc}), and that obtained from \textcolor{black}{exact} dynamics $T^\text{typ}_\text{eq}$ for different observables and $\lambda$ ($\lambda \in [0.1, 2.0]$ for ${\cal A}_{\pi}$ and ${\cal A}_{\pi/12}$, and $\lambda \in [0.1, 4.0]$ for ${\cal A}_\Delta$). (a)-(d) refer to the corresponding
cases illustrated in Figs.~\ref{Fig1} and \ref{Fig2}.
 }\label{Fig3}
\end{figure}

If $\delta_S$ is small ($\delta_S < 0.2$), as shown by open dots in Fig.~\ref{Fig3}, we observe $T^\text{rc}_\text{eq}\approx T^\text{typ}_\text{eq}$, which indicates the high accuracy of our estimation in this regime.
In case of larger $\delta_S$ ($\delta_S \ge 0.2$),   $T^\text{rc}_\text{eq}$ begins to deviate from  $T^\text{typ}_\text{eq}$.
This suggests that once the Lanczos coefficients are smooth, our method tends to converge with good accuracy. For more detailed data on the relation of smoothness of Lanczos coefficients and the and the accuracy of our approach see Fig.~S6 in the supplemental material. \new{In the End Matter, we examine the spin density wave operator in the Heisenberg XXZ model and also find that our approach yields good accuracy when the $b_n$ grow in a smooth fashion.}

Based on these findings  we believe our methods  does converge for a very wide range of autocorrelation functions in fully chaotic systems as a multitude of numerical studies reveal the respective Lanczos coefficients $b_n$  to be smooth early on, see e.g.\ \cite{Parker19,PhysRevE.106.014152-Bn-Robin22,PhysRevB.109.L140301-Bn-Oleg,wang2023diffusion-RM-wang24}.
It appears that smoothness of Lanzcos coefficients and chaos are interlinked. For the average gap ratios $\langle r\rangle$\cite{Oganesyan-PhysRevB.75.155111} corresponding to the Hamiltonians of our four main examples, illustrated in Figs.~\ref{Fig1} and \ref{Fig2} we find (for $L = 18$) (a): $\langle r\rangle=0.53$, (b)(c): $\langle r\rangle=0.53$, (d): $\langle r\rangle=0.46$. Note that full chaoticity is characterized by $\langle r\rangle\approx0.53$. Hence the case with poor convergence, (d), is also the least chaotic case.

{\it Conclusion and Outlook.\ }
In this paper, based on the recursion method, we suggest a scheme to estimate 
the equilibration times of local observables, by making use of the corresponding Lanczos coefficients. 
We numerically find that such estimations converge quickly when the first several Lanczos coefficients are taken into account, provided that the observable eventually features smoothly growing Lanczos coefficients. It implies that the first several Lanczos coefficients are sufficient for a reasonable estimate of $T_\text{eq}$ in the thermodynamic limit.
Our numerical findings are further supported by analytical arguments.

\new{A natural future direction is to apply our approach to other types of systems, including the Fermi Hubbard model, models with disorder as well as higher dimensional systems. It is also interesting to extend our framework to time-dependent systems, or, Floquet systems, to estimate the equilibration time towards time-dependent steady states.
Benchmarking the results against those from large-scale systems using Density Matrix Renormalization Group (DMRG) calculations could also provide important insights to further assess the accuracy of the approach.}

{\it Outline of supplemental material.\ }
In the supplemental material we argue that for smooth Lanczos coefficients the formula $B_N\approx2b_{N/2}$ for the Lanczos coefficients of the squared dynamics holds approximately. Based on the observation that the Krylov vectors $\vert Q_n)$ subject to the scheme in Eq.\ (\ref{eq-Bn}) are primarily located at the outmost counterdiagonal of the product space $\{\vert n)\otimes\vert m)\}$, the argument is laid out in the framework of a stochastic process imposed by $\mathcal{L}$.

{\it Acknowledgement.\ }
This work has been funded by the Deutsche
Forschungsgemeinschaft (DFG), under Grant No. 531128043, as well as under Grant
No.\ 397107022, No.\ 397067869, and No.\ 397082825 within the DFG Research
Unit FOR 2692, under Grant No.\ 355031190. Additionally, we greatly acknowledge computing time on the HPC3 at the University of Osnabr\"{u}ck, granted by the DFG, under Grant No. 456666331.

\bibliographystyle{apsrev4-1_titles.bst}
\bibliography{main.bib}

\clearpage

\setcounter{section}{0}
\setcounter{secnumdepth}{2}

\begin{figure}[!htbp]
	\centering
    \includegraphics[width = 1.0\linewidth]{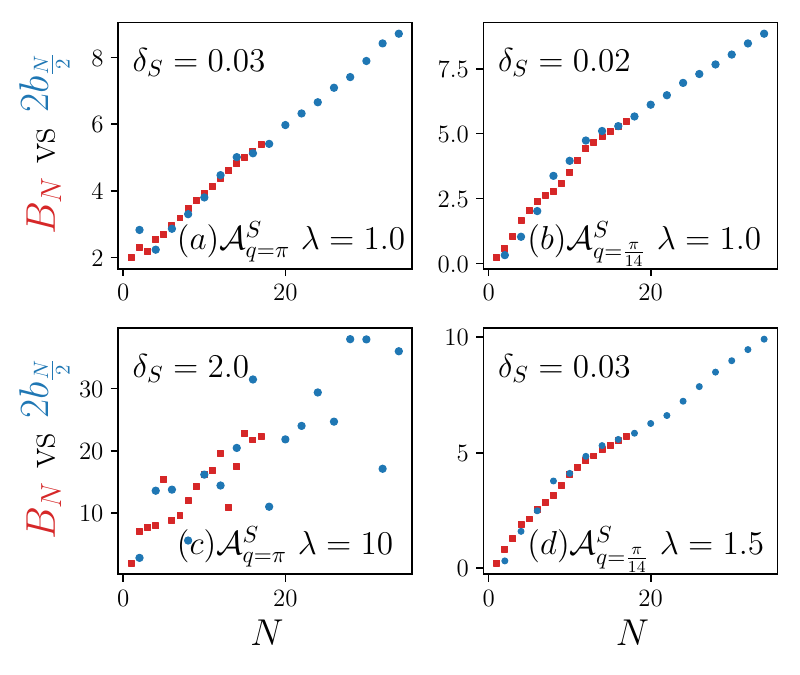}
	\caption{Lanczos coefficients $B_N$ (red square) related to the autocorrelation function $\mathcal{C}^2$ versus $2 b_{N/2}$ (blue dot) with $b_n$ the Lanczos coefficients related to $\mathcal{C}$, for the spin density wave operators in XXZ model. The results are shown for different $\lambda$ and wave number $q$. 
    (a): $q = \pi,\ \lambda = 1.0$; (b) $q = \pi/14,\ \lambda = 1.0$; (c): $q = \pi,\ \lambda = 10.0$ and (d): $q = \pi/14,\ \lambda = 1.5$.
    $\delta_S$ is the smoothness indicator defined in Eq.~\eqref{eq-smooth}.
    }\label{Fig4}
\end{figure}

\begin{figure}[!htbp]
	\centering
    \includegraphics[width = 1.0\linewidth]{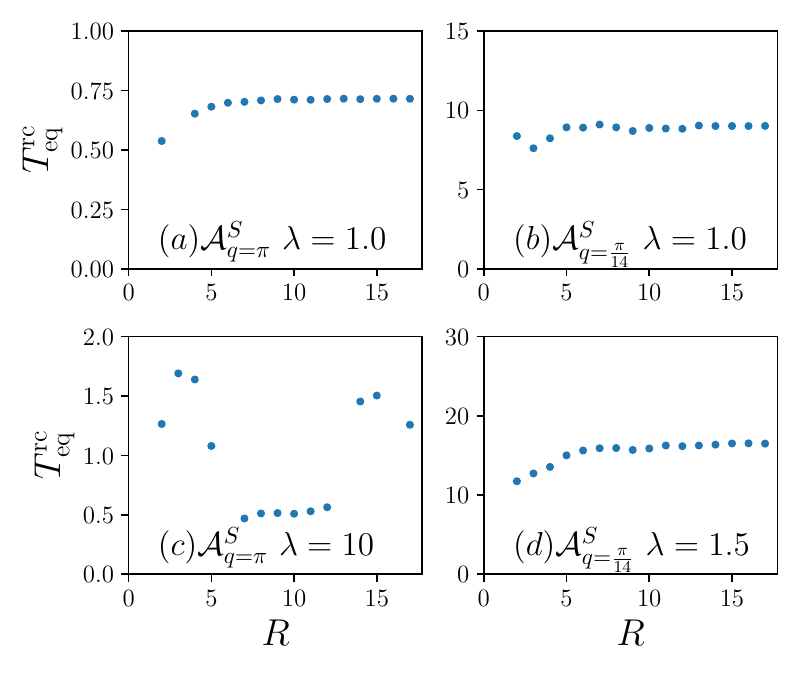}
	\caption{Equilibration time $T^{\text{rc}}_{\text{eq}}$ (given in Eq.~\eqref{eq-T-rc}) for different numbers $R$ of included Lanczos coefficients, for the spin density wave operators in XXZ model. The results are shown for different $\lambda$ and wave number $q$. 
    (a): $q = \pi,\ \lambda = 1.0$; (b) $q = \pi/14,\ \lambda = 1.0$; (c): $q = \pi,\ \lambda = 10.0$ and (d): $q = \pi/14,\ \lambda = 1.5$. 
}\label{Fig5}
\end{figure}

\begin{figure}[!htbp]
	\centering
    \includegraphics[width = 0.9\linewidth]{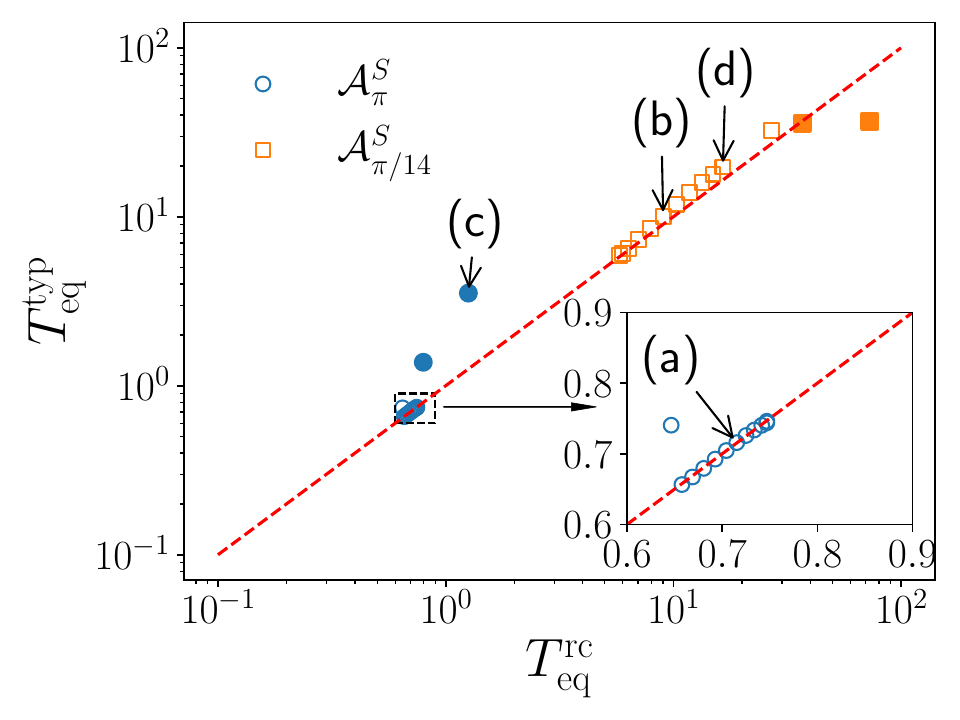}
	\caption{Comparison between the equilibration time estimated using the recursion method $T^\text{rc}_\text{eq}$ (Eq.~\eqref{eq-T-rc}), and that obtained from exact dynamics $T^\text{typ}_\text{eq}$ for
    the spin density wave operator ${\cal A}^S_q$ in XXZ model with wave number $q=\pi, \pi/14$ and $\lambda$ ($\lambda \in [0.5, 10.0]$.
    Open (solid) dots indicates results for $\delta_S < 0.2$ ($\delta_S \ge 0.2$). Here we choose $R=n_{\text{max}}.$ The inset shows a zoomed-in region for better distinguishability of the results.
 }\label{Fig6}
\end{figure}

\begin{center}
    \textbf{End Matter}
\end{center}
In the End Matter, we present numerical results for spin density wave operators in the Heisenberg XXZ model.
The Hamiltonian reads 
\begin{equation}
   H=\sum_{\ell=1}^{L}J(s_{\ell}^{x}s_{\ell+1}^{x}+s_{\ell}^{y}s_{\ell+1}^{y}+\lambda s_{\ell}^{z}s_{\ell+1}^{z}+\Delta^{\prime}s_{\ell}^{z}s_{\ell+2}^{z}),
\end{equation}
where $s_{\ell}^{x,y,z}=\frac{1}{2}\sigma_{\ell}^{x,y,z}$ indicates the spin operator at site $\ell$. Parameters are chosen as $J=1.0, \Delta^\prime = 0.5$ . 
Here we consider the spin density wave operator
\begin{equation}
    {\cal A}_{q}^{S}=\sum_{\ell=1}^{L}\cos(q\ell)s_{\ell}^{z},
\end{equation}
where $q$ indicates the wave number and we focus on $q = \pi$ and $q = \pi/14$. The results are shown in Figs.~\ref{Fig4},~\ref{Fig5} and \ref{Fig6}, analogue to Figs.~\ref{Fig1},~\ref{Fig2} and \ref{Fig3} in the main text. For both observables we obtain $n_{\text{max}} = 17$ Lanczos coefficients.

In Fig.~\ref{Fig4}, we present some examples for Lanczos coefficients, where $B_N \approx 2 b_{\frac{N}{2}}$ is observed in cases where the original $b_n$ grows in a smooth fashion (indicated by a small $\delta_S$), as shown in [(a)(b)(d)]. In contrast, if the $b_n$ is not smooth, its non-smooth structure carries over to $B_N$, leading to deviations between 
$B_N$ and $2 b_{\frac{N}{2}}$. This is similar to the finding in Fig.~1.

In Fig.\ \ref{Fig5} we show our estimation of $T_{\text{eq}}^{\text{rc}}$ for varying $R$ for the cases [(a)-(d)] discussed in Fig.\ \ref{Fig4}. Similar to the observables shown in Fig.~\ref{Fig2},  if Lanczos coefficients are smooth (as in the case of [(a)(b)(d)]), $T_{\text{eq}}^{\text{rc}}$ only varies slightly with respect to $R$, for $R\gtrsim5$. In contrast, 
when the Lanczos coefficients exhibits significant fluctuations (in the case of (c)), the estimate $T_{\text{eq}}^{\text{rc}}$ does not converge, indicating that an accurate result for $T_{\text{eq}}$ might not yet have been reached.

Furthermore, in Fig.~\ref{Fig6} we benchmark the accuracy of our estimate $T^{\text{rc}}_{\text{eq}}$ against the results from the direct simulations (using DQT) of the autocorrelation functions, denoted by $T^{\text{typ}}_{\text{eq}}$. In the DQT simulations, we consider system size $L = 28$.
Good agreement between $T^{\text{rc}}_{\text{eq}}$ and $T^{\text{typ}}_{\text{eq}}$ is observed if $\delta_S$ is small (open dots), suggesting the high accuracy of our estimation.
For large values of $\delta_S$ ($\delta_S \ge 0.2$), the deviation of $T^\text{rc}_\text{eq}$ from $T^\text{typ}_\text{eq}$ becomes noticeable (see, for example, point (c)).

To summarize, the additional numerics on the spin density wave operator in the Heisenberg model further suggest the result of the main text that, if the observable eventually shows smoothly  growing Lanczos coefficients, a finite number of the former is sufficient for a reasonable estimate of the equilibration time.

\clearpage
\newpage

\setcounter{figure}{0}
\setcounter{equation}{0}
\renewcommand{\thefigure}{S\arabic{figure}}
\renewcommand{\theequation}{S\arabic{equation}}

\section*{Supplemental Material}

\subsection*{Outline}
Here we provide some analytical arguments that firstly, the concentration of the Krylov vector along the forefront of product space $\{\vert n_1)\otimes \vert n_2)\}$ is a self-consistent assumption, putting forward analytical evidence that for pertinent classes of Lanczos coefficients the evolution of the Krylov vector is particularly simple. Secondly, we translate the successive action of the Liouvillian into stochastic process. This enables us to infer statements on the structure of the Krylov vectors associated to the Lanczos coefficients $B_n$ of the squared autocorrelation function $\mathcal{C}^2(t)$. Lastly, we bring together both of the above statements formulate an approximate relation between the Lanczos coefficients $B_n$ of the squared dynamics and those of the original autocorrelation function, $b_n$.

Furthermore we provide 1): our estimate $T_{\text{eq}}^{\text{rc}}$ for the toy model with smooth Lanczos coefficients $b_n=\sqrt{n}$; 2): numerical results on the time integral of squared auto-correlation functions and 3): detailed derivations of Eq.~(14) in the main text.

\subsection*{Concentration on forefront}
\begin{figure}[t]
	\centering
    \includegraphics[width = 1.0\linewidth]{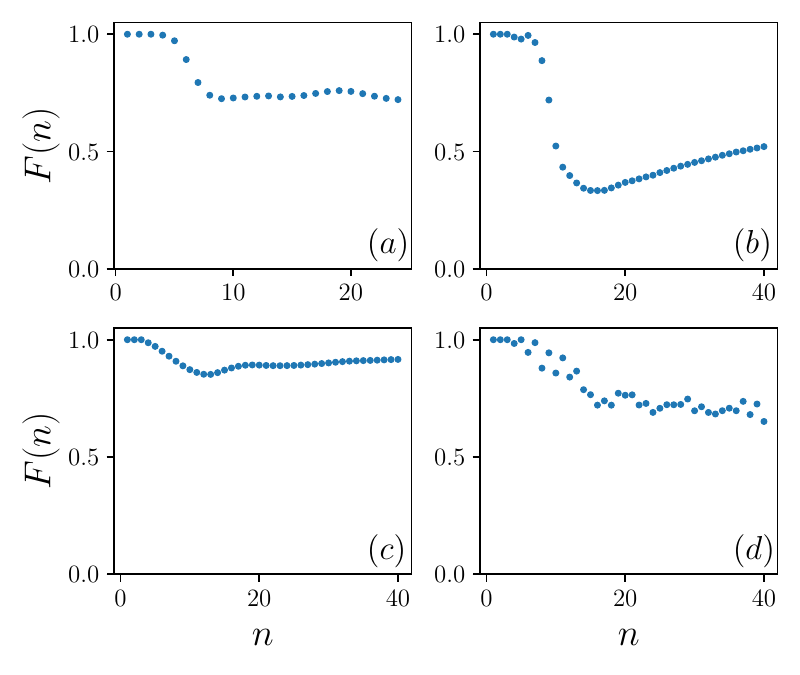}
	\caption{\textit{Concentration on the forefront}:
    $F(n)$ (Eq.~\eqref{eq-Fa}) versus $n$
    for (a): Ising ladder, $\mathcal{A}_\Delta$, $\lambda=0.5$; (b): TFI, $\mathcal{A}_q (q=\pi/12)$, $\lambda=0.5$; (c): TFI, $\mathcal{A}_q(q=\pi)$, $\lambda=0.5$ and (d): TFI, $\mathcal{A}_q(q=\pi)$, $\lambda=2.0$. 
    }\label{Fig-Falpha}
\end{figure}

Numerically we find that for pertinent Lanczos coefficients the Krylov vector for the squared dynamics, i.e.\ generated by $\mathcal{L}=\mathcal{L}_1+\mathcal{L}_2$, for smooth Lanczos coefficients, is predominantly located at outmost counterdiagonal, see Fig.\ \ref{Fig-Falpha} for a first impression. We refer to this behaviour as \textit{concentration on the forefront}. 
First, we introduce the labeling of states $\vert \alpha-\beta)\otimes\vert\beta)=:\vert\alpha,\beta)$. Any state on the product space may be written as
\begin{align}
    \vert n)=\sum_{\alpha,\beta=0}\Phi^n(\alpha,\beta)\vert\alpha,\beta).
\end{align}

We assume that $\Phi^n(\alpha,\beta)\approx0$ for $\alpha\neq n$, i.e.\ that the vector is concentrated along the outmost counterdiagonal, the \textit{forefront}. We check the self-consistency of this assumption, testing the appropriateness of the assumption in the first place. To this end, we formally rewrite action of the Liouvillian onto a state in two parts that raise (lower) in the value of $\alpha$, i.e.\ propagate the state the next-higher (next-lower) counterdiagonal,
\begin{align}
    (\mathcal{L}_1+\mathcal{L}_2)\vert n)&=:\mathcal{L}^-\vert n)+\mathcal{L}^+ \vert n),\\
    (\alpha,\beta\vert \mathcal{L}^-\vert n)&=0\quad\text{for}\quad\alpha\neq n-1,\label{eq_lowering}\\
    (\alpha,\beta\vert \mathcal{L}^+\vert n)&=0\quad\text{for}\quad\alpha\neq n+1.\label{eq_raising}
\end{align}

\begin{figure}[h]
 	\centering
     \includegraphics[width = 0.7\linewidth]{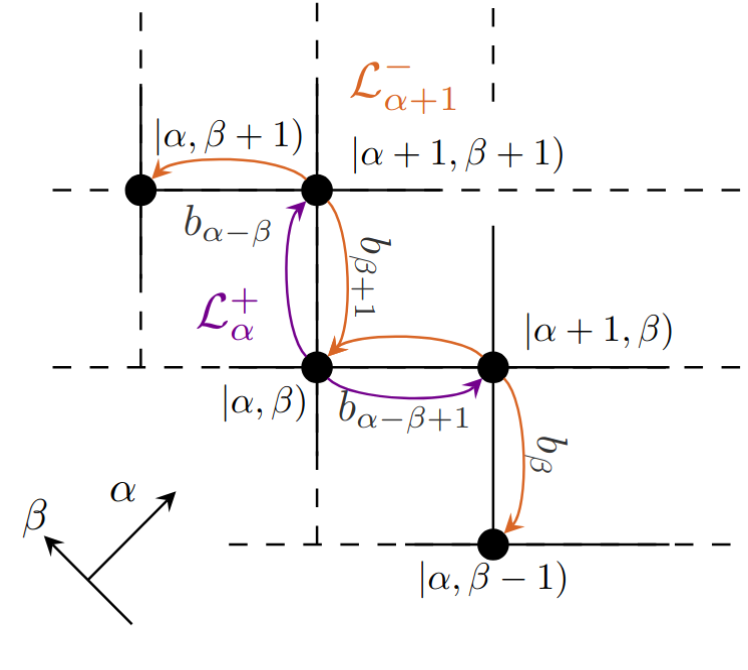} 	\caption{Sketch of the action of $\mathcal{L}^-\mathcal{L}^+ \vert n)$ on the product grid spanned by the Krylov state $\{\vert n_1)\otimes\vert n_2)\}$, ``zoomed in" on the state $\vert\alpha,\beta)$. \label{Fig-L_PM}}
 \end{figure}

\begin{figure}[t]
	\centering
    \includegraphics[width = 1.0\linewidth]{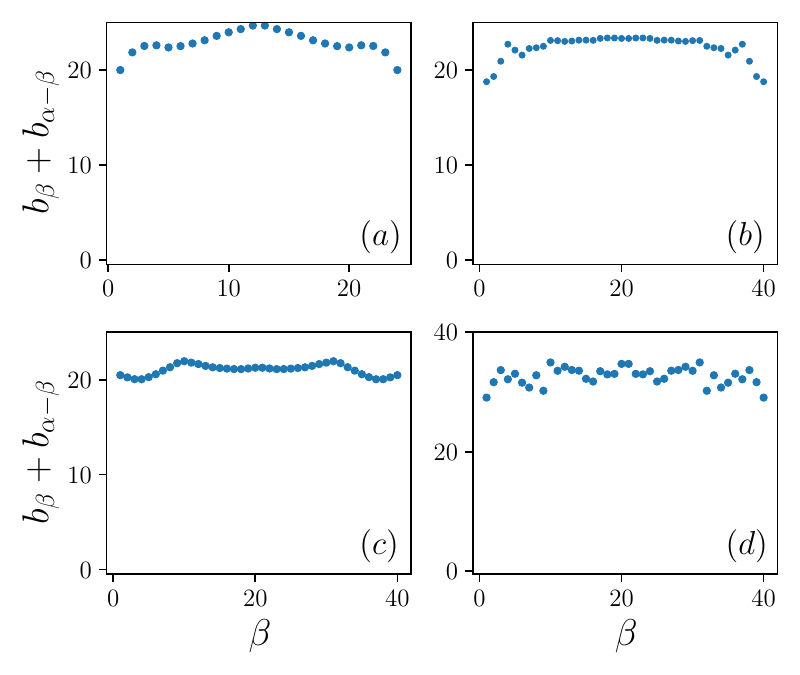}
	\caption{$b_{\beta}+b_{\alpha-\beta}$ versus $\beta$ for (a): Ising ladder, $\mathcal{A}_\Delta$, $\lambda=0.5$; (b): TFI, $\mathcal{A}_q (q=\pi/12)$, $\lambda=0.5$; (c): TFI, $\mathcal{A}_q(q=\pi)$, $\lambda=0.5$ and (d): TFI, $\mathcal{A}_q(q=\pi)$, $\lambda=2.0$. Here we choose $\alpha = n_{\text{max}}$.
    }\label{FigS5}
\end{figure}

The first Lanczos step finds
\begin{align*}
\begin{split}
        \vert n+1)&\propto \mathcal{L}^-\vert n)+\mathcal{L}^+\vert n)\\&-\left[( n-1\vert \mathcal{L}^-\vert n)+(n-1\vert \mathcal{L}^+ \vert n)\right]\vert n-1).
\end{split}
\end{align*}
By virtue of Eq. (\ref{eq_raising}) the last term vanishes.
Hence, to stay in line with the initial assumption we need to have $\vert n-1)\propto \mathcal{L}^-\vert n)$.
Next, we turn to the second Lanczos step
\begin{align}
  \begin{split}
        \vert n+2)&\propto \mathcal{L}^- \mathcal{L}^+ \vert n)+(\mathcal{L}^+)^2\vert n)\\&-\left[(n\vert \mathcal{L}^-\mathcal{L}^+\vert n)+(n\vert(\mathcal{L}^+)^2\vert n)\right]\vert n).
  \end{split}
\end{align}
As before, the last term vanishes due to Eq. (\ref{eq_raising}) and in order to find $\vert n+2)\propto(\mathcal{L}^+)^2\vert n)$ we need to have $\mathcal{L}^-\mathcal{L}^+\vert n)\propto\vert n)$. However, in contrast to the first Lanczos step, we may check this explicitly. Starting from a specific state $\vert\alpha,\beta)$ the successive action of $\mathcal{L}^-\mathcal{L}^+$ involves 5 states and 4 Lanczos coefficients in total (one less resp. at the edges), as illustrated in Fig.\ \ref{Fig-L_PM}. First, we define $\chi(\beta):=(\alpha+1,\beta\vert \mathcal{L}^+\vert n)$ as the amplitude of the \textit{raised} state along the counterdiagonal $\alpha+1$.
For this, in total 4 terms, originating from 3 different states along $\alpha$ contribute, see Fig.\ \ref{Fig-L_PM}.
Concretely, we have
\begin{align}
\chi(\beta)&=b_\beta\Phi^n(\alpha,\beta-1)+b_{\alpha-\beta+1}\Phi^n(\alpha,\beta),\\
    \chi(\beta+1)&=b_{\alpha-\beta}\Phi^n(\alpha,\beta+1)+b_{\beta+1}\Phi^n(\alpha,\beta).
\end{align}
Finally, the action of $\mathcal{L}^-$ maps the state back to the  counterdiagonal $\alpha$. Denoting the overlap along this initial counterdiagonal by $\Psi(\alpha,\beta):=(\alpha,\beta\vert \mathcal{L}^-\mathcal{L}^+\vert n)$ we find
\begin{align}
    &\Psi(\alpha,\beta)=-\left(b_{\beta+1}\chi(\beta+1)+b_{\alpha-\beta+1}\chi(\beta)\right)\\
    \begin{split}
        &=-\left(b_{\beta+1}b_{\alpha-\beta}\Phi^n(\alpha,\beta+1)+b^2_{\beta+1}\Phi^n(\alpha,\beta)\right.\\
  & \left.b_{\alpha-\beta+1}b_{\beta}\Phi^n(\alpha,\beta-1)+b^2_{\alpha-\beta+1}\Phi^n(\alpha,\beta) \right).\label{eq-forefront}
    \end{split}
\end{align}
If we further assume that the Lanczos coefficients change little, i.e.\ $b_n\approx b_{n+1}$, and that along some counterdiagonal $\alpha$ the profile is reasonably smooth, i.e.\ $\Phi^n(\alpha,\beta-1)\approx\Phi^n(\alpha,\beta)\approx\Phi^n(\alpha,\beta+1)$ Eq. (\ref{eq-forefront}) simplifies to
\begin{align}
    \Psi(\alpha,\beta)\approx-\left(b_\beta+b_{\alpha-\beta}\right)^2\Phi^n(\alpha,\beta).
\end{align}
From this we can infer that  if
\begin{align}
    b_\beta+b_{\alpha-\beta}=\text{const.}\quad\text{with respect to $\beta$},\label{eq-condition}
\end{align}
applies, $\vert n+2)\propto(\mathcal{L}^+)^2\vert n)$ may be fulfilled, hence concentration on the forefront may occur, see Fig.\ \ref{FigS5} for a check of Eq.\ (\ref{eq-condition}) for the physical cases (a)-(d). This also allows or a corresponding conclusion in an approximate sense. However, the condition (\ref{eq-condition}) is not only strictly fulfilled in the scenario of purely linear Lanczos coefficients $b_n$ but also for Lanczos coefficients of the form $b_n=a n+c$.  It may also be approximately fulfilled for other sets of  $b_n$. Revisiting Fig.\ \ref{Fig-Falpha}, as well as Fig.\ 1 in the main text, we find that in the cases of suitably linear (and hence smooth) Lanczos coefficients (see cases (a)-(c)), i.e.\ cases that approximately fulfill the condition (\ref{eq-condition}), there is a concentration on the forefront. Conversely, in (d) the $b_n$ deviate considerably from a linear form in the numerically accessible regime and the Krylov vector $\vert Q_n)$ is hardly located at the forefront.   
We check the assumption of concentration on the forefront by considering 
\begin{equation}\label{eq-Fa}
F(n)=\sum_{\beta=0}^{\alpha}|\Phi^{n}(\alpha=n,\beta)|^{2},
\end{equation}
and the results are shown in Fig.~\ref{Fig-Falpha}.

\subsection*{Stochastic matrices}

\begin{figure}[h]
 	\centering
     \includegraphics[width = 1.0\linewidth]{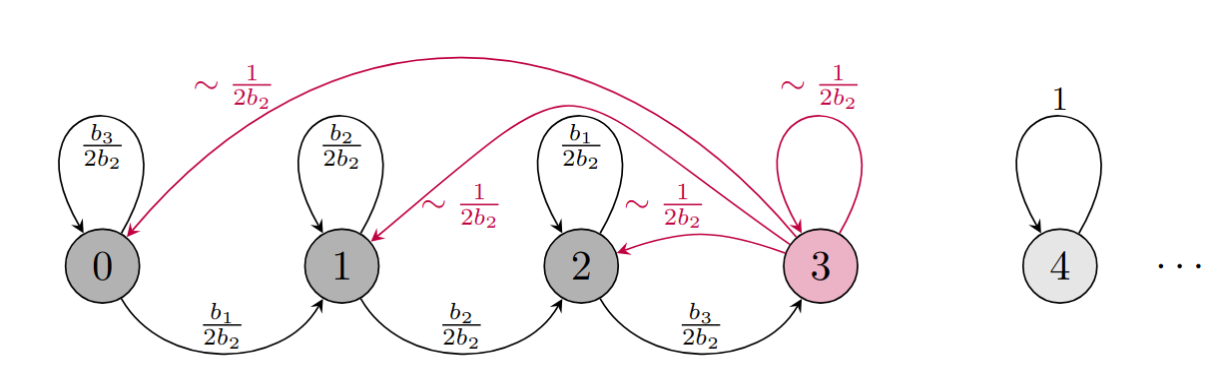} 	\caption{Markov chain of the stochastic process corresponding to the action of $\mathcal{M}_2$. The input here is given by the three entries $p^2_0, p^2_1,p^2_2$ of the occupation vector on the second counterdiagonal, indicated by the gray circles ``0",``1",``2". The entry at ``3" is initially zero, i.e.\ $p^2_3=0$, hence the contribution to the other states indicated by the arrow $\sim\frac{1}{2b_2}$ have no weight in the stochastic process. However, there is are non-zero rate towards ``3", such that $p^3_3\neq0$ for the subsequent step of the process, i.e.\ $\mathcal{M}_3$ (not shown). \label{Fig-Markov}}
 \end{figure}

Our aim is to translate the repeated action of $\mathcal{L}^+$ into a stochastic process. Assuming an absolute concentration on the forefront, we consider a normalised state $\vert Q_n)=\sum_{k=0}^n p^n_k\vert n,k)$ that is solely contained along the $n^{\text{th}}$ counterdiagonal of the product space spanned by $\{\vert n_1)\otimes\vert n_2)\}$.
Consequently, we focus on sets of Lanczos coefficients $b_n$ that (locally) fulfill condition (\ref{eq-condition}), i.e.\ those that are (locally) of the form $b_n=an+c$.
The operator $\mathcal{L}_{n}^+$ corresponds to the action of the Liouvillian propagating the amplitudes of a state from the counterdiagonal $n$ to the next-highest counterdiagonal $n+1$. Concretely,
\begin{align}
\mathcal{L}_n^+=\begin{pmatrix}
       b_n&0&0&\dots&0\\
       b_1&b_{n-1}&0&\dots&0\\
       0&b_2&\ddots&\ddots&0\\
       \vdots&\ddots&\ddots&\ddots&\vdots\\
       \vdots&\ddots&\ddots&b_{n-1}&b_1\\
       0&\dots&\dots &0&b_n
    \end{pmatrix}\label{eq_lplus}.
\end{align}
For better clarity, we turn to square matrices
and consider operators of the form
\begin{equation}
\mathcal{M}_{n,d} =
\begin{pmatrix}
\frac{1}{2\left(an/2+c\right)}
\begin{bmatrix}
\mathcal{L}^+_n & \Bigg\vert & 
\begin{matrix}  
v_0 \\ 
\vdots \\ 
v_n 
\end{matrix}
\end{bmatrix} & 0 \\ 
0 & \mathbb{1}_{d-n}
\end{pmatrix},
\end{equation}
where $v_0=v_n=a+c$ and else $v_k=a$ for $a$ and $c$ from the linear form of the $b_n$. Here, several remarks are in order. First, the dimension-padding imposed by the identity matrix is done such that for some counterdiagonal $d$ all matrices have the same dimension. For clarity of notation, we refrain from carrying the subscript explicitly, i.e.\ $\mathcal{M}_n\hat{=}\mathcal{M}_{n,d}.$
Secondly, the matrix $\mathcal{M}_n$ is \textit{doubly-stochastic}, i.e.\ $\sum_j (\mathcal{M}_n)_{jk}=\sum_k (\mathcal{M}_n)_{jk}=1$.
Further, we understand the occupation vectors as being padded into the proper ambient dimension, e.g.\ along the counterdiagonal $n$ the occupation vector reads:
\begin{align}
    p^n=(p^n_0,p^n_1,\dots,p^n_n,\underbrace{0,\dots,0}_{d-n}).
\end{align}
With this it becomes evident that when computing \textit{new} occupation amplitudes of the state at the counterdiagonal $n+1$ we have
\begin{align}
    p^{n+1}&=\mathcal{M}_{n+1}\ p^n,\\
    &=\Pi_{j=2}^{n+1}\mathcal{M}_{j}\ p^1
\end{align}
for which neither for the dimensionality-padding constructions for $\mathcal{L}^+_n$, $\mathcal{M}_n$ or $p^n$ enter. For better accessibility, we show in Fig.\ \ref{Fig-Markov} the respective Markov chain for the transition from counterdiagonal $n=2\rightarrow3$, i.e.\ the action of $\mathcal{M}_2$.

For every counterdiagonal $n$ we consider the quantity:
\begin{align}
    \mathrm{H}^n=-\sum_j p_j^n\ln\left(\frac{p_j^n}{q_j^n}\right),\label{eq_H-function1}
\end{align}
where the $\{q_i^n\}$ denote equilibrium probabilities of the stochastic process generated by the $\mathcal{M}_n$ and the $p_j^n$ label the entries of the occupation vector $p^n$, i.e.\ the actual probabilities at the position $j$ at the iteration $n$. By virtue of Ref.\ \cite{penrose-foundations-book} the quantity defined by Eq.\ (\ref{eq_H-function1}) is monotonous with respect to $n$ and hence identifies the direction of the approach to equilibrium of the stochastic process imposed by $\mathcal{L}$.

For each $\mathcal{M}_n$ we have $q_j^n\equiv1/d$ which follows from the double stochasticity of the $\mathcal{M}_n$, hence
\begin{align}
    \mathrm{H}^n=-\sum_j p_j^n\left(\ln p_j^n-\ln d\right).\label{eq-htilde}
\end{align}
constitutes a monotonous function in $n$. 
This quantity (up to the constant addend $-\ln d$) may be interpreted as the (increasing) Shannon entropy of the stochastic process along $n$. Consequently, the growing entropy implies the successive \textit{smoothening} of the occupation vectors $p^n_j$ along $n$ and with respect to $j$, i.e.\ the distribution of the Krylov states $\vert Q_n)$ gets successively wider. 
\begin{figure}[t]
	\centering
    \includegraphics[width = 1.0\linewidth]{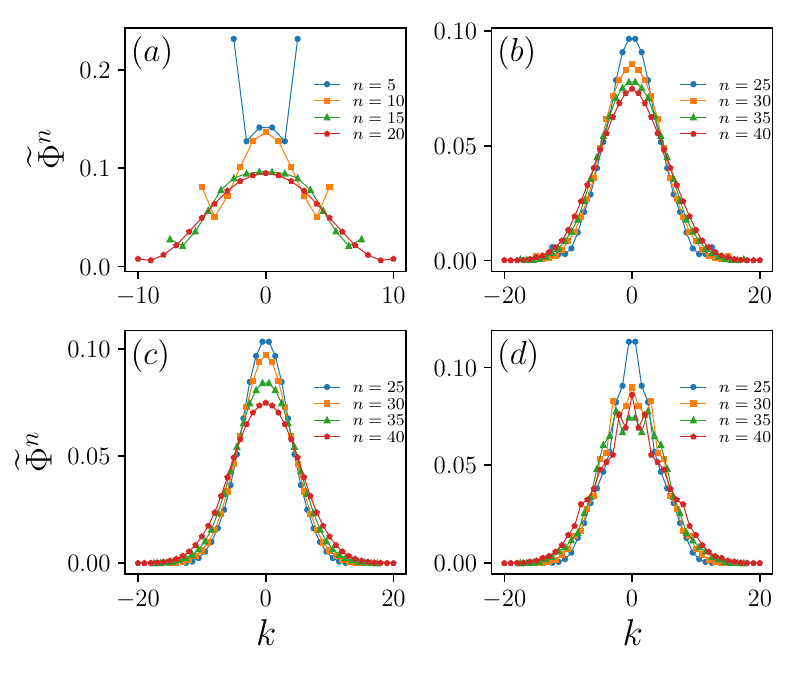}
	\caption{\textit{Distribution along the forefront}: Spread of the normalized coefficients $\tilde{\Phi}^n(\alpha=n,k)\propto p^n_k$ at several counterdiagonals $n$ for
    (a): Ising ladder, $\mathcal{A}_\Delta$, $\lambda=0.5$; (b): TFI, $\mathcal{A}_q (q=\pi/12)$, $\lambda=0.5$; (c): TFI, $\mathcal{A}_q(q=\pi)$, $\lambda=0.5$ and (d): TFI, $\mathcal{A}_q(q=\pi)$, $\lambda=2.0$.
    }\label{FigS4}
\end{figure}
Note that in the setting of the preceding subsection this finding translates to an increasing applicability  of the assumption $\Phi^n(\alpha,\beta-1)\approx \Phi^n(\alpha,\beta)\approx \Phi^n(\alpha,\beta+1)$.

For Lanczos coefficients of operators in real quantum many-body systems, the picture is clearly more intricate than in the scenario of linear Lanczos coefficients above, as the matrices $\mathcal{M}_n$ are generally not fully \textit{doubly stochastic}. However, also in the physical cases (b) and (c) studied throughout this paper we find that the corresponding Lanczos coefficients $b_n$ \textit{locally} fulfill condition (\ref{eq-condition}), see Fig.\ \ref{FigS5}, whereas for case (d) the condition is violated. Here we leave out case (a), as in this scenario the accessible number of Lanczos coefficients is too low for an adequate analysis. However, the plateau that becomes visible in the bulk, suggests that eventually also in this case condition (\ref{eq-condition}) will be fulfilled.

In Fig.\ \ref{FigS4} we examine the distribution of the Krylov vectors $\vert Q_n)$ at several forefronts $n$, i.e.\ the spread of the coefficients $\Phi^n(\alpha=n,k)$ along $k$. Relating to the language of occupation vectors and amplitudes,  this is equivalent to the spreading of the occupation amplitudes $p^n_k$ along $k$ for various $n$. We find that in the cases of smooth Lanczos coefficients, that (locally) comply with (\ref{eq-condition}), the occupation amplitudes spread out, reminiscent of a diffusive process (here (a)-(c)). Again the case (d) stands out as the profile does not smoothen out, contrary to the other scenarios in which consequently the idea of a growing (Shannon) entropy, see Eq.\ (\ref{eq-htilde}), becomes tangible.



\begin{figure}[h]
	\centering
    \includegraphics[width = 1\linewidth]{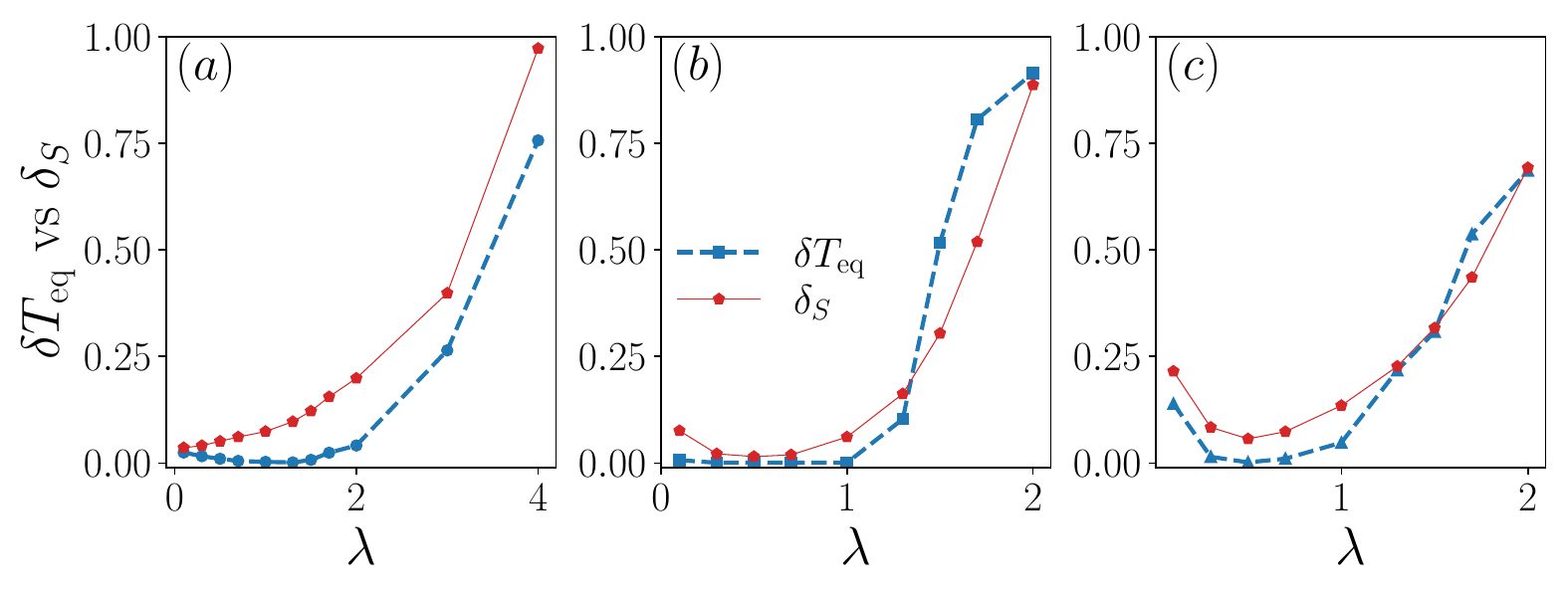}
	\caption{$\delta T_{\mathrm{eq}}=\frac{|T_{\text{eq}}^{\text{rc}}-T_{\text{eq}}^{\text{typ}}|}{T_{\text{eq}}^{\text{typ}}}$ (solid line) and smoothness $\delta_S$ (dashed line) versus $\lambda$ for operators (a): Ising ladder ${\cal A}_\Delta$, (b) TFI: ${\cal A}_{\pi}$ and (c) ${\cal A}_{\pi/12}$.
 }\label{Fig-S1}
\end{figure}

\begin{figure}[]
	\centering
    \includegraphics[width = 0.7\linewidth]{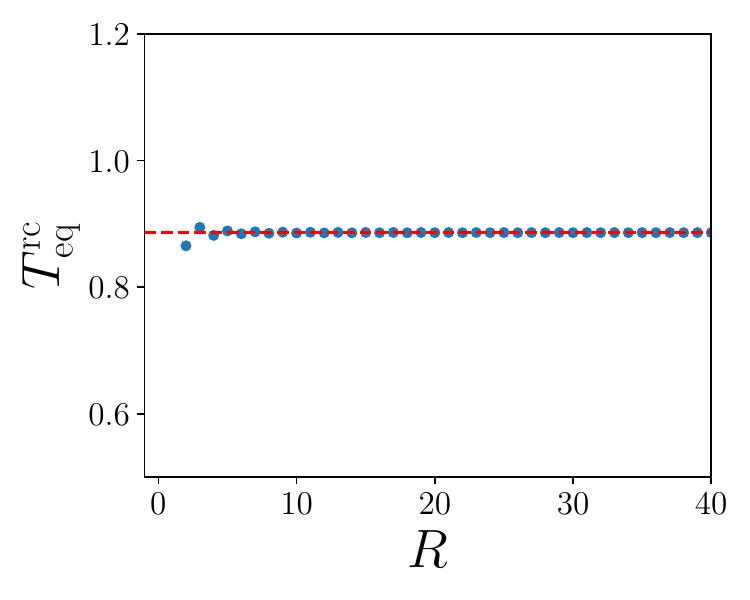}
	\caption{$T^{\text{rc}}_{\text{eq}}$  as a function of $R$ (starting point of the linear continuation of $B_N$), for
 the toy model $b_n = \sqrt{n}$. The dashed line indicates the analytical prediction $T_\text{eq} = \frac{\sqrt{\pi}}{2}$.  
    }\label{FigS2}
\end{figure}

\subsection*{Approximation formula}
Consider the amplification of a normalized vector $\vert Q_n)$ under the action of $\mathcal{L}^+_n$. Assuming a full concentration on the forefront, the $j^{\text{th}}$ entry of the vector $\vert\Tilde{Q}_{n+1}):=\mathcal{L}^+_n\vert Q_n)$ gets amplified by a factor $b_{n-j}+b_j$. Note that for the case of $ b_k = a k +c$ we get
\begin{equation}
    b_{n-j} + b_j = an+2c = 2 b_{\frac{n}{2}}
\end{equation}

For the Lanczos coefficient related to the this new Krylov vector we have 
\begin{align}
    B_{n+1}=\sqrt{(\Tilde{Q}_{n+1}\vert\Tilde{Q}_{n+1})}\label{eq-bn-q}.
\end{align}
If we assume additionally that the vector $\vert Q_n)$ is spread out sufficiently even (see preceding section), we may infer an approximation of the $B_n$ by
\begin{align}
    B_n\approx 2b_{\frac{n}{2}}. \label{eq-approx}
\end{align}

\subsection*{Benchmark cases}
One of the few examples of analytically known connections between autocorrelation functions and Lanczos coefficients is given by 
\begin{align}
    C(t)=\exp\left(-\frac{t^2}{2}\right)\longleftrightarrow b_n=\sqrt{n}.
\end{align}
This case is especially interesting as it also allows to infer the Lanczos coefficients for the squared autocorrelation function, which here also turns out to be a Gaussian (with different variance),
\begin{align}
    C^2(t)=\exp\left(-t^2\right)\longleftrightarrow B_n=\sqrt{2n}.
\end{align}
We find that in this case $2b_{n/2}=\sqrt{2n}=B_n$, and hence that our approximation formula (\ref{eq-approx}), derived on the basis of a stochastic process, is exact.

For purely linear Lanczos coefficients $b_n=\alpha n$ the Lanczos algorithm on the product space $\{\vert n_1)\otimes\vert n_2)\}$ is particularly simple as the contributions by $\mathcal{L}^-$ vanish in the orthogonalization and the Krylov vector is hence fully located at the forefront. For this special case the distribution along the counterdiagonal is constant, as the number of "paths" to each site on the counterdiagonal and the corresponding "weight" given by the Lanczos coefficients cancel,
\begin{align}
    \vert Q_n)=\frac{1}{\sqrt{n+1}}\sum_{k=0}^n\vert n-k)\vert k).
\end{align}
From this the Lanczos coefficients of the squared autocorrelation function follow as
\begin{align}
    B_n=\alpha\sqrt{n(n+1)}.
\end{align}
In this scenario, for large $n\gg1$ we find linear growth and therefore an approximate agreement with Eq.\ (\ref{eq-approx}).
In fact, the case of entirely linear Lanczos coefficients is special as in this case the spread of the Krylov vectors $\vert Q_n)$ is uniform. When computing the Lanczos coefficient $B_n$ as in Eq.\ (\ref{eq-bn-q}) we must take into account, that the dimension of the $\vert Q_{n-1})$ is $n$ whereas $\vert Q_n)=\mathcal{L}^+_n\vert Q_{n-1})/\Vert\cdot\Vert$ is $n+1$. With this we find
\begin{align}
    B_n=2b_{n/2}\sqrt{\frac{n+1}{n}}=\alpha\sqrt{n(n+1)}.
\end{align}
However, in general the influence from the growing dimension is negligible, since the spread of the Krylov vectors falls off towards the edges, see Fig.\ \ref{FigS4}. Therefore we generally expect the simple approximation (\ref{eq-approx}) to hold well.

\begin{figure}[]
	\centering
    \includegraphics[width = 1.0\linewidth]{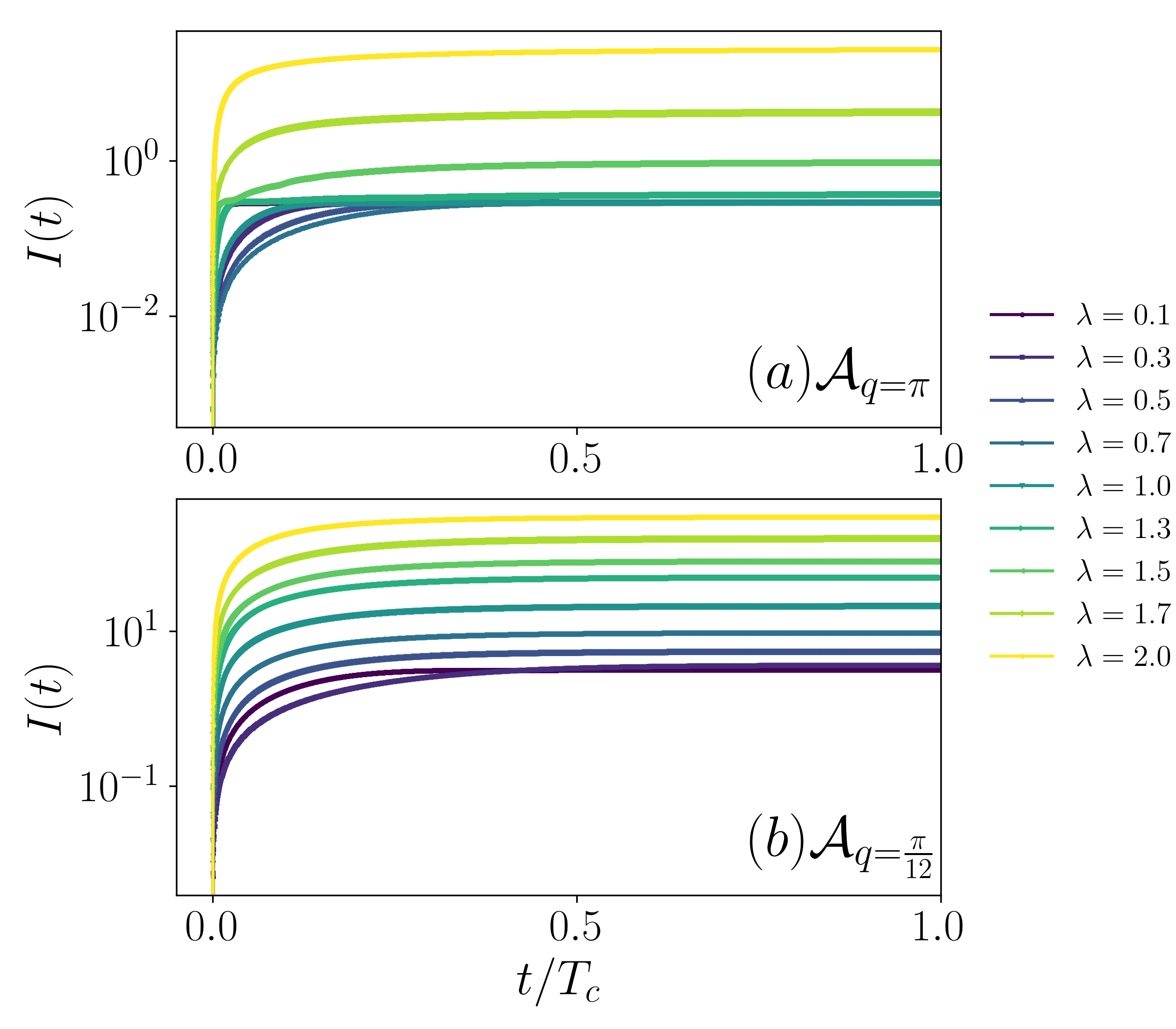}
	\caption{Time integral of the squared autocorrelation function: $I(t)$ (defined in Eq.~\eqref{eq-It}) versus $t/T_c$ for (a): ${\cal A}_{q=\pi}$ and (b): ${\cal A}_{q=\frac{\pi}{12}}$ in the tilted field Ising model for different $\lambda$. The system size is $L = 24$.
    }\label{Fig-Dt-Ising}
\end{figure}

\begin{figure}[]
	\centering
    \includegraphics[width = 1.0\linewidth]{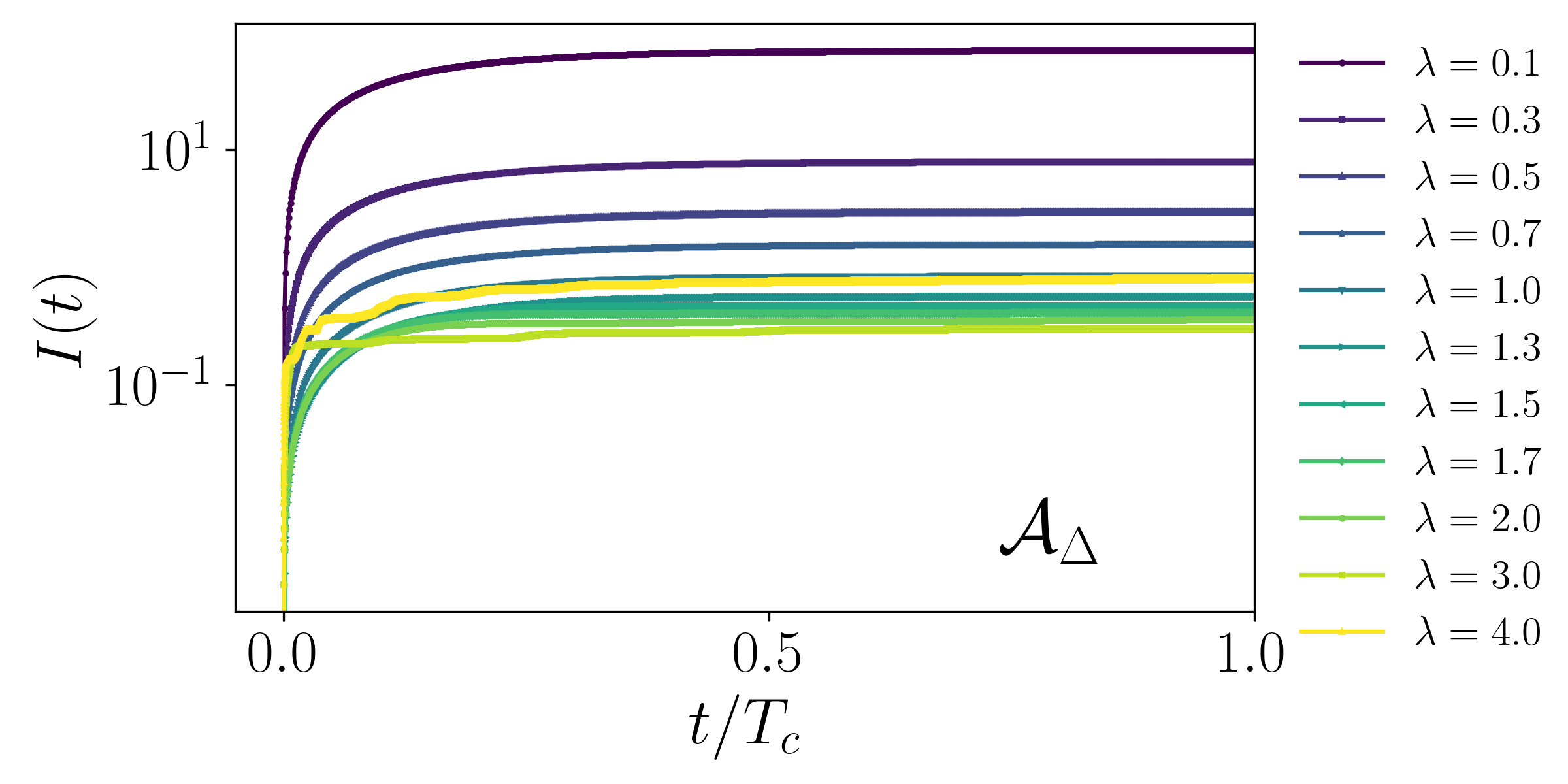}
	\caption{Similar to Fig.~\ref{Fig-Dt-Ising}, but for ${\cal A}_{\Delta}$ in the Ising ladder. The system size is $L = 24$.  
    }\label{Fig-Dt-Wheel}
\end{figure}

\begin{figure}[]
	\centering
    \includegraphics[width = 1.0\linewidth]{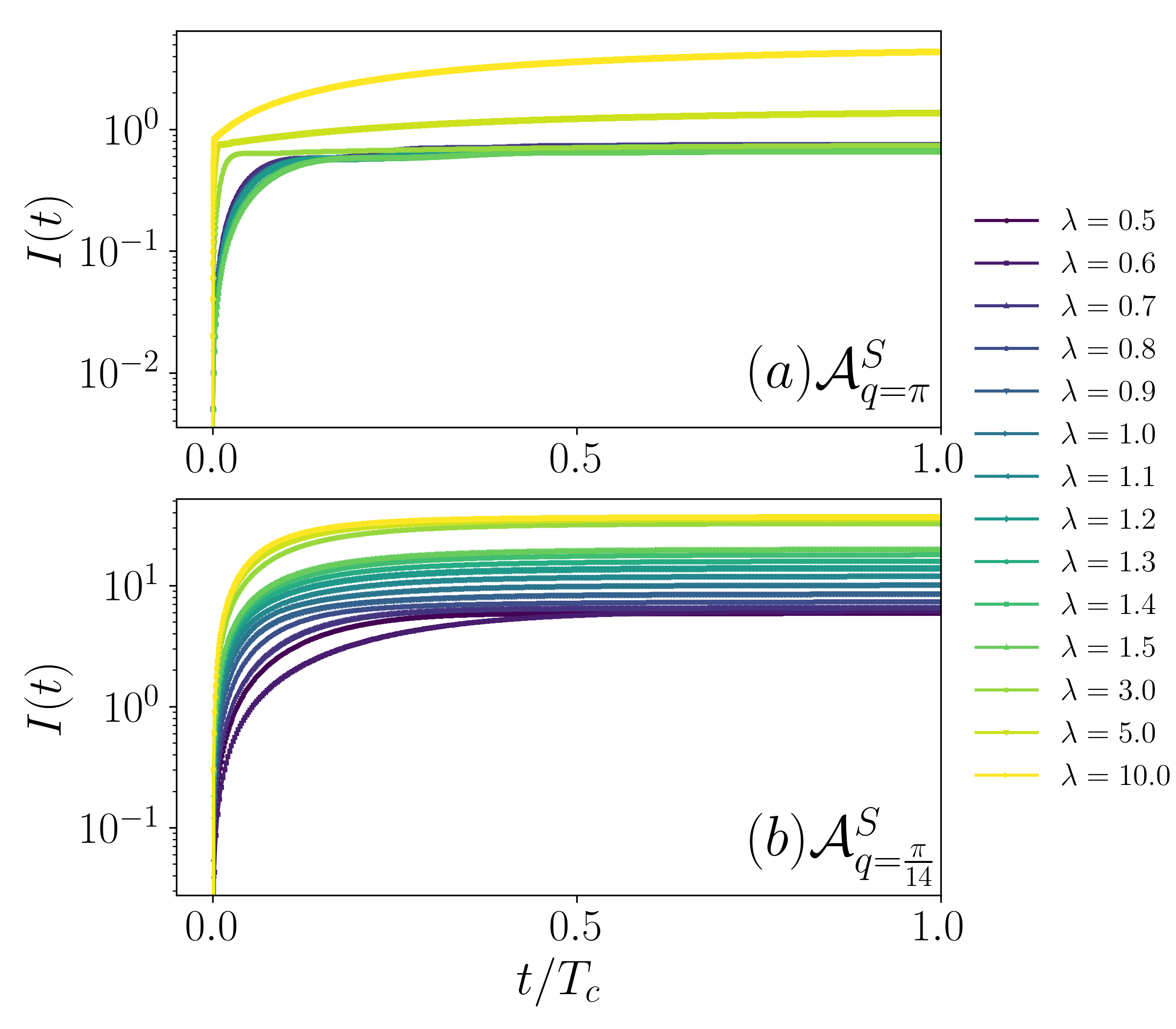}
	\caption{Similar to Fig.~\ref{Fig-Dt-Ising}, but for (a):${\cal A}^S_{q=\pi}$ and (b): ${\cal A}^S_{q=\frac{\pi}{14}}$ in the XXZ model. The system size is $L = 28$.
    }\label{Fig-Dt-XXZ}
\end{figure}

\subsection*{Derivation of Eq.~(14) in the main text}
In this section, we show the detailed derivation of Eq.~(14) in the main text.
To this end, we insert $B_N(N>R) = \alpha N + \beta$ to Eq.~(15). This leads to
\begin{gather}
p_{R}=\prod_{k=1}^{\infty}\frac{(B_{2k+R-1})^{2}}{B_{2k+R}b_{2k-2+R}}\nonumber\\
=\prod_{k=1}^{\infty}\frac{\left(\alpha(2k+R-1)+\beta\right)^{2}}{\left(\alpha(2k+R)+\beta\right)\left(\alpha(2k-2+R)+\beta\right)} \nonumber \\
=\prod_{k=1}^{\infty}\frac{\left(k+\frac{\alpha R+\beta}{2\alpha}-\frac{1}{2}\right)^{2}}{\left(k+\frac{\alpha R+\beta}{2\alpha}\right)\left(k+\frac{\alpha R+\beta}{2\alpha}-1\right)}. \label{eq-pn1}
\end{gather}
Making use the following expression of Gamma function 
\begin{equation}
\Gamma(z+1)=\prod_{k=1}^{\infty}\left[\frac{1}{1+\frac{z}{k}}\left(1+\frac{1}{k}\right)^{z}\right],
\end{equation} 
it is easy to get 
\begin{equation}\label{eq-GammaP}
\frac{\Gamma(z-a+1)\Gamma(z+a+1)}{\Gamma^{2}(z+1)}=\prod_{k=1}^{\infty}\frac{(k+z)^{2}}{(k+z-a)(k+z+a)}.
\end{equation}
Comparing Eq. \eqref{eq-GammaP} with Eq. \eqref{eq-pn1},
and setting $z=\frac{R}{2}+\frac{\beta}{2\alpha}-\frac{1}{2},\quad a = \frac{1}{2}$, one has
\begin{equation}
p_{R}=\frac{\Gamma(\frac{R}{2}+\frac{\beta}{2\alpha})\Gamma(\frac{R}{2}+\frac{\beta}{2\alpha}+1)}{\Gamma^{2}(\frac{R}{2}+\frac{\beta}{2\alpha}+\frac{1}{2})},
\end{equation}
which is the result of Eq.~(14).

\section{More details on the calculation of $T^{\text{typ}}_{\text{eq}}$}
$T^{\text{typ}}_{\text{eq}}$ is defined as
\begin{equation}
T_{\text{eq}}^{\text{typ}}:=\int_{0}^{T_{c}}|{\cal C}_{\text{typ}}(t)|^{2}dt,
\end{equation}
where ${\cal C}_{\text{typ}}(t)$ represents the autocorrelation function calculated via dynamical quantum typicality\ \cite{PhysRevLett.102.110403-Christian-DQT} (DQT):
\begin{equation}
    {\cal C}_{\text{typ}}(t)=\langle\psi|{\cal O}(t){\cal O}|\psi\rangle.
\end{equation}
The state $|\psi\rangle$ is a normalized Haar-random state, the real and imaginary part of which are Gaussian random numbers. 
$T_c$ indicates the cutoff time, which is determined by the time  after which $|{\cal C}_\text{typ}(t)|^2\le0.001$ for all computable times $t\ge T_c$. To further check whether $T_{\text{eq}}^{\text{typ}}$ is insensitive to the cutoff time $T_c$, in Figs.~\ref{Fig-Dt-Ising},~\ref{Fig-Dt-Wheel} and ~\ref{Fig-Dt-XXZ}, we plot
\begin{equation}\label{eq-It}
    I(t) = \int _0 ^t |{\cal C}_{\text{typ}}(s)|^2ds
\end{equation}
in the tilted field Ising model, Ising ladder and XXZ model and various of observables. Note that $T^\text{typ}_{\text{eq}} = I(t=T_c)$.
It is clear from the figures that, for all cases considered, $I(t)$ approximately saturate at $t< T_c$. This implies that $T_{\text{eq}}^{\text{typ}}$ is indeed insensitive to the cutoff time $T_c$.

\end{document}